# Energy intake functions of ectotherms and endotherms derived from their body mass growth

Authors

Jan Werner[1,4], Nikolaos Sfakianakis[2,4], Alan Rendall[3] and Eva Maria Griebeler[1]

[1] Evolutionary Ecology, Zoological Institute, Johannes Gutenberg-University Mainz, Germany

[2] Analyses, Mathematical Institute, Johannes Gutenberg-University Mainz, Germany

[3] Numeric, Mathematical Institute, Johannes Gutenberg-University Mainz, Germany

[4] contributed equally to the manuscript

Corresponding author

Name: Jan Werner

Mailing address: Department of Ecology, Zoological Institute, University of Mainz, P.O. Box 3980, D-55099 Mainz, Germany

Phone: +49 (0)6131-39 22718

Fax: +49(0)6131-39 23731

E-mail: wernerja@uni-mainz.de

# Abstract

How animals allocate energy to different body functions is still not completely understood and a challenging topic until recently. Here, we investigate in more detail the allocation of energy intake to growth, reproduction or heat production by developing energy budget models for ectothermic and endothermic vertebrates using a mathematical approach. We calculated energy intake functions of ectotherms and endotherms derived from their body mass growth. We show that our energy budget model produces energy intake patterns and distributions as observed in ectothermic and endothermic species. Our results comply consistently with some empirical studies that in endothermic species, like birds and mammals, energy is used for heat production instead of growth. Our model additionally offers an explanation on known differences in absolute energy intake between ectothermic fish and reptiles and endothermic birds and mammals. From a mathematical point of view, the model comes in two equivalent formulations, a differential and an integral one. It is derived from a discrete level approach, and it is shown to be well-posed and to attain a unique solution for (almost) every parameter set. Numerically, the integral formulation of the model is considered as an inverse problem with unknown parameters that are estimated using a series of experiments/realistic data.

# 1 Introduction

## 1.1 Background

Every organism needs energy from its biotic and/or abiotic environment to maintain its body functions, to grow and to reproduce. How animals allocate energy to these different tasks is still not completely understood and a challenging topic until recently (Hou et al., 2008; Kooijman, 2010; Sousa et al., 2008; West et al., 2001). Only the endothermic vertebrates (birds and mammals) use surplus energy to maintain their body temperature, whereas in ectothermic vertebrates (fish and reptiles) the body temperature is mainly set by ambient temperatures and not by internal physiological heat production as in endotherms (Clarke and Pörtner, 2010; Cowles, 1940; Grigg et al., 2004; Koteja, 2004; McNab, 1978; Nespolo et al., 2011; Schweitzer and Marshall, 2001). Endothermy in birds and mammals is enabled by a basal metabolic rate which is high enough to permit high and constant body temperatures (within a few degrees Celsius) within the whole body and under a broad range of ambient temperatures (Clarke and Pörtner, 2010; Cowles, 1940). Thus, birds and mammals have the advantage that their metabolic rate and energy intake does not so much depend on the environmental temperature as in ectotherms, but compared to ectotherms they have the disadvantage that they need additional energy to fuel internal heat production. As a result, endothermic birds and mammals have an increased metabolic rate and energy intake compared to similar-sized ectotherms, even if both have similar body temperatures (Brown et al., 2004; Hulbert and Else, 1981; Ruben, 1995; White et al., 2006). However, endothermic birds and mammals are not endothermic from the beginning of their life (McClure and Randolph, 1980; Ricklefs, 1979; Ricklefs, 1987; Visser and Ricklefs, 1995). Even after hatching or birth, in most birds and mammals endothermy is not fully developed (McClure and Randolph, 1980; Ricklefs, 1979; Ricklefs, 1987; Visser and Ricklefs, 1995). Studies on growth and development of endothermic organisms suggested a trade-off between growth and endothermy (McClure and Randolph, 1980; Węgrzyn, 2013). Energy allocated to growth is not available for heat production and vice versa. As it is experimentally very difficult to assess energy allocation in living organisms, especially over their whole life, instead theoretical models, so-called energy budget models, have been developed to address this issue. In the 1990's, Kozlowski and colleagues did important work in developing optimization models, which aimed at optimal age to allocate energy to growth or to reproduction (Kozlowski, 1992; Kozlowski, 1996; Kozlowski and Weiner, 1997). Kooijman (2010) developed a complex dynamic energy budget (DEB) theory based on a few biological assumptions, which can describe energy budgets of organisms within space and time. Here, we investigate in more detail the allocation of energy intake to growth, reproduction or heat production by developing energy budget models for ectothermic and endothermic vertebrates. Our model might be seen as a modification and extension of Kozlowski's (1992) energy budget model and a special case of Kooijman's (2010), dynamic energy budget (DEB) theory, as it primarily focuses on the two different thermal regulation strategies seen in vertebrates. In particular, we assume that ectothermic and endothermic individuals allocate their energy intake to maintenance, growth, and reproduction, whereas only endotherms allocate energy to heat production, too. We also assume that the fractions of energy allocated to growth, reproduction and heat production change over ontogenetic time and the life of the animal. Using empirical growth data, we show that our energy budget model produces energy intake patterns and distributions as observed in ectothermic and endothermic species.

## 1.2 Paper Overview

This paper starts with a mathematical description of the proposed energy budget model and with an investigation of its mathematical consistency and well-posedness (2). We then bring the mathematical formulations into computational numerical descriptions (3). Then, we describe the implementation and parameterisation of our energy budget model (4). At the end of the paper, we present model results and discuss their biological implications (5).

## 2 The energy budget model

In this section, we establish an energy budget model based on mathematical formulations. It describes a possible allocation of relative energy to maintenance, growth, reproduction, and in the case of endothermy to heat production of a vertebrate over its lifespan. In combination with an absolute energy intake function it describes the growth of an individual within its life. We show that the model is mathematically consistent, well-posed and behaves in a biologically reasonable way.

In particular, we assume that a modelled individual represents the average of its species and that the individual lives in an environment providing enough resources to fulfil all its energetic needs for maintenance, growth, reproduction and heat production (100% of its energy intake = 100% of its energetic needs). Intra- and interspecific competition between animals for resources/ energy is absent.

## 2.1 Energy allocation to maintenance, growth, reproduction and heat production

The total energy intake of an individual is given by $E_{intake}(t)$ where $t \in [0, t_{max}]$ is the life-time variable of the individual and $t_{max}$ the maximal longevity observed within the species. $E_{intake}(t)$ is allocated to *growth* $E_{grow}(t)$ and *maintenance* $E_{maint}(t)$ of the body, to *heat production* $E_{heat}(t)$, and to *reproduction* $E_{repr}(t)$ as:

$$E_{intake}(t) = E_{maint}(t) + E_{grow}(t) + E_{heat}(t) + E_{repr}(t). \tag{1}$$

Accordingly, we define the corresponding *relative energies* with respect to the total energy intake $E_{intake}$ as:

$$\bar{E}_{maint}(t) = \frac{E_{maint}(t)}{E_{intake}(t)}, \quad \bar{E}_{heat}(t) = \frac{E_{heat}(t)}{E_{intake}(t)},$$

$$\bar{E}_{grow}(t) = \frac{E_{grow}(t)}{E_{intake}(t)}, \quad \bar{E}_{repr}(t) = \frac{E_{repr}(t)}{E_{intake}(t)}. \tag{2}$$

We further make the assumption that the relative maintenance energy remains constant through the life of the individual $\bar{E}_{maint}(t) = \bar{E}_{maint}$; this *does not* imply that the *absolute* maintenance energy $E_{maint}$ remains constant. Furthermore, we make, for the relative heat and reproduction energies $\bar{E}_{heat}, \bar{E}_{repr}$ the following biologically reasonable assumption (see McClure and Randolph, 1980; Ricklefs, 1979; Ricklefs, 1987; Visser and Ricklefs, 1995): an initial phase of no (or irrelevant small) energy investment is followed by a phase of linear increase and by a stagnation phase during ontogeny:

$$\bar{E}_{heat}(t) = \begin{cases} 0, & 0 \le t \le t_h^1 \\ \text{linear}, & t_h^1 < t \le t_h^2, \\ \bar{E}_{heat}^{max}, & t_h^2 < t \le 1 \end{cases} \quad (3)$$

$$\bar{E}_{repr}(t) = \begin{cases} 0, & 0 \le t \le t_r^1 \\ \text{linear}, & t_r^1 < t \le t_r^2, \\ \bar{E}_{repr}^{max}, & t_r^2 < t \le 1 \end{cases} \quad (4)$$

where $t_h^1, t_h^2, t_r^1, t_r^2 \left(= \frac{t}{t_{max}}\right) \in [0,1]$ are *relative-time* instances that signify the changes in the nature of $\bar{E}_{heat}, \bar{E}_{repr}$.

We allow the growth energy to be a "free variable" given as the complement of the other three relative energies, i.e.

$$\bar{E}_{grow} = 1 - \bar{E}_{maint} - \bar{E}_{repr} - \bar{E}_{heat}. \quad (5)$$

With the above assumptions, and for a given energy intake function, we deduce that the relative energy distribution is uniquely determined by the seven, species dependent, parameters (Figure 1)

$$\{\bar{E}_{maint}, \bar{E}_{heat}^{max}, \bar{E}_{repr}^{max}, t_h^1, t_h^2, t_r^1, t_r^2\}. \quad (6)$$

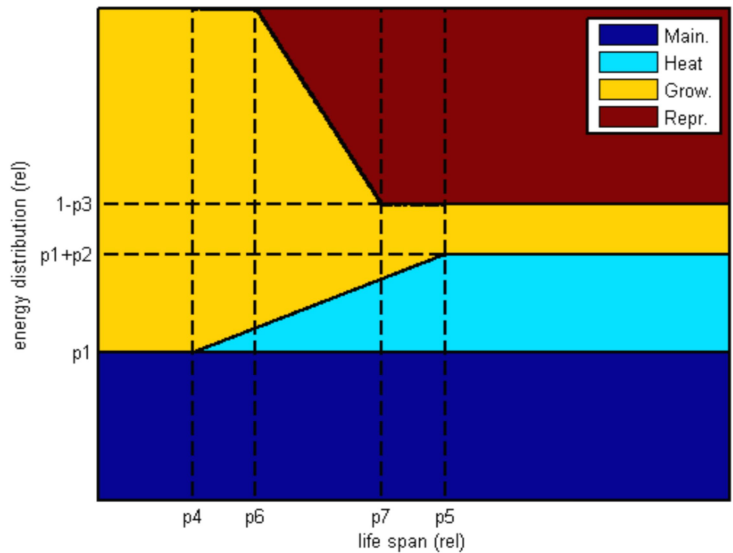

**Figure 1.** A graphical representation of the relative energy distribution over the lifespan of an arbitrary individual. The parameters $\{p1, p2, \cdots, p7\}$ correspond to the parameter set (6): p1 = $\bar{E}_{maint}$, p2 = $\bar{E}_{heat}^{max}$, p3 = $\bar{E}_{repr}^{max}$; p4 = $t_h^1$; p5 = $t_h^2$, p6 = $t_r^1$; p7 = $t_r^2$. In the particular case of an ectothermic species we expect p2=0.

Only for an ectothermic species, we expect $\bar{E}_{hea}^{max} = 0$.

## 2.2 Absolute energy intake and body mass growth of an individual

We propose in this section a mathematical model that relates the body mass growth of an animal as a function of its life-time to its absolute energy intake, and to the absolute growth $E_{grow}(\cdot)$ and maintenance energies $E_{maint}(\cdot)$.

Our model is based on the premises that the growth and the maintenance of the body are *quasi-independent* processes. They both depend on the energy intake $E_{\text{intake}}$, and although the growth term can only dictate an increase in body mass, the lack of sufficient maintenance energy might lead it to decrease. The final result is a combination of these two, possibly contradictory, effects.

The proposed model reads in its *differential formulation* as follows:

$$m'(t) = \bar{E}_{maint} E'_{intake}(t) + \bar{E}_{grow}(t) E_{intake}(t). \tag{7}$$

where $m'(t)$ represents the rate of change in body mass, and the two terms in the right hand side describe the contributions to the body mass by the maintenance and the growth energy respectively. We comment briefly here on these terms and refer to the Section 9 and to (8) for the derivation of (7).

On the one hand, the growth term in the right-hand side of (7) indicates that a part of the total energy intake $E_{\text{intake}}$ is invested in growth unless otherwise "dictated by the animal", that is unless $\bar{E}_{grow} = 0$. Since though $E_{intake} \geq 0$ and $\bar{E}_{grow} \geq 0$ the growth term describes, exclusively, *an increase* in the body mass.

Considering on the other hand only the maintenance term we see that an increase in body mass ($m'(t) \geq 0$) necessitates an increase of the energy needed for maintenance. This in turn implies, since the relative maintenance energy $\bar{E}_{maint}$ is assumed to be constant, an increase in the total energy intake ($E'_{intake} \geq 0$). Conversely, the model also predicts that a decrease of the total energy intake ($E'_{intake} \leq 0$), due to e.g. lack of resources, leads to a reduction in body mass ($m'(t) \leq 0$).

As we previously noted, both efforts of maintenance and growth work in parallel and independently from each other. This can be seen in the extreme case of *no energy intake* where the model predicts *some loss of mass* but *not all of it*; the mass that has been accumulated will mostly remain. The same can happen (conditionally) in the case of reduced energy intake. In both cases, the outcome is in agreement with our understanding of the development of the body mass when the resources are limited.

The proposed model (7) can also be given in a somewhat different formulation that is easier to handle numerically

$$m(t) = E_{maint}(t) + \int_0^t E_{grow}(\tau)d\tau. \tag{8}$$

The *integral formulation* (8) can be justified by a discrete approach, see Section 9. The differential version (7) is directly derived from (8) using the additional assumption of the relative maintenance energy $\bar{E}_{maint}$ being constant. In particular, by employing (2) we can reformulate (8) as:

$$m(t) = \bar{E}_{maint}(t)E_{intake}(t) + \int_0^t \bar{E}_{grow}(\tau)E_{intake}(\tau)d\tau. \tag{9}$$

After taking the time derivative, (9) reads as

$$m'(t) = \bar{E}'_{maint}(t)E_{intake}(t) + \bar{E}_{maint}(t)E'_{intake}(t) + \bar{E}_{grow}(t)E_{intake}(t),$$

which recasts to (7), after recalling that the relative maintenance energy $\bar{E}_{maint}$ is constant.

## 2.3 Backward approach (well-posedness of the model)

We have so far considered the case where the energy intake, as well as the energy distribution, is given, and we evaluate the development of the body mass over time. It is also instructive and biologically challenging to investigate the backward approach, where the development of the body mass $m'(t)$ is given (empirical data on growth are frequently compared to information on energy intake and its allocation to tasks) and (7) is considered as an Ordinary Differential Equation (ODE) with respect to the energy intake $E_{intake}$. It turns out that we can solve this ODE easily:

First multiply (7) by $e^{\int_0^t \frac{\bar{E}_{grow}(s)}{\bar{E}_{maint}} ds}$ and after integration over time obtain

$$E_{intake}(t) = e^{-\int_0^t \frac{\bar{E}_{grow}(s)}{\bar{E}_{maint}} ds} \int_0^t \frac{m'(s)}{\bar{E}_{maint}} e^{\int_0^s \frac{\bar{E}_{grow}(\tau)}{\bar{E}_{maint}} d\tau} ds,$$

which can also be written as

$$E_{intake}(t) = \int_0^t \frac{m'(s)}{\bar{E}_{maint}} e^{\int_0^s \frac{\bar{E}_{grow}(\tau)}{\bar{E}_{maint}} d\tau - \int_0^t \frac{\bar{E}_{grow}(s)}{\bar{E}_{maint}} ds} ds, \qquad (10)$$

or

$$E_{intake}(t) = \int_0^t \frac{m'(s)}{\bar{E}_{maint}} e^{\int_t^s \frac{\bar{E}_{grow}(\tau)}{\bar{E}_{maint}} d\tau} ds. \qquad (11)$$

In the more general case where $\bar{E}_{maint}(\cdot)$ is *not a constant*, the corresponding differential version of (9) reads:

$$m'(t) = \bar{E}_{maint}(t) E'_{intake}(t) + \left(\bar{E}'_{maint}(t) + \bar{E}_{grow}(t)\right) E_{intake}(t), \qquad (12)$$

which can then be solved for $E_{intake}$ to give:

$$E_{intake}(t) = \int_0^t \frac{m'(s)}{\bar{E}_{maint}(s)} e^{\int_t^s \frac{\bar{E}'_{maint}(\tau) + \bar{E}_{grow}(\tau)}{\bar{E}_{maint}(\tau)} d\tau} ds. \qquad (13)$$

The benefit of considering (11) and/or (13) is mostly analytical and not computational since they provide the energy intake function, given that the body mass growth and the energy distribution are known. Practically this means that the relative energy distributions, determined in our model by the seven coefficients (6) are needed before employing (11) and/or (13).

## 3 Implementation of the energy budget model (Numerical treatment)

To numerically treat (8), we first consider a discretization $M_t = \{0 = t_1, \cdots, t_{N+1} = t_{max}\}$ of the species longevity $[0, t_{max}]$ into $N$ equivalent *computational cells* $C_k = [t_k, t_{k+1}]$, where $t_1 = 0$ and $t_{k+1} - t_k = \Delta t = \frac{t_{max}}{N}$ for all $k = 1, \cdots, N$.

For the *forward problem*, i.e. the numerical computation of the body mass from the energy intake and distribution over an animal's life, we discretize (9) over $M_t$ in the sense of *Finite Volumes* as:

$$m^n = \bar{E}_{maint} E^n_{intake} + \sum_{k=1}^{n} \bar{E}^k_{grow} E^k_{intake} \Delta t \tag{14}$$

where $m^k, E^k_{intake}, \bar{E}^k_{grow}$ are numerical approximations of the mass, the absolute energy intake, and the relative growth energies respectively evaluated at $C_k$.

### 3.1 Backward approach

Our overall objective is to *identify* the energy intake and energy distribution of an animal from its (*experimentally observed*) growth curve. In more detail, we consider the evolution of the body mass $m^{exp}$ as given, and identify a (computational) body mass curve $m^{com}$ that approximates $m^{exp}$ "adequately". The computational $m^{com}$ follows from a) the energy intake $E_{intake}$, b) the energy distribution as given by the parameters (6), see also Figure 1, and c) the numerical version (14) of the model (9). The notion of "adequate" approximation is quantified below in (17). To this end, we moreover assume (as a first guess) that the absolute energy intake follows a *logistic function* of the life of the animal:

$$E_{intake}(t) = \frac{M}{1 + be^{-kt}}, \quad b = \frac{M}{m_0} - 1. \tag{15}$$

4The additional parameters $\{m_0, M, k\}$ introduced as a result of this assumption are also subject to identification.

In summary, given an experimental body growth curve $m^{exp}$ we need to estimate the ten parameters,

$$\{\bar{E}_{main}, \bar{E}_{heat}^{max}, \bar{E}_{repr}^{max}, \tau_h^1, \tau_h^2, \tau_r^1, \tau_r^2, m_0, M, k\} \quad (16)$$

so that the relative *root mean square* (RMS) difference between $m^{exp}$ and $m^{com}$ (henceforth RMS error) is less than a predefined threshold $0 < \varepsilon < 1$:

$$E_{rel} = \frac{\|m^{exp} - m^{com}\|_{RMS}}{\|m^{exp}\|_{RMS}} < \varepsilon \quad (17)$$

It should be noted that $m^{com}$ is derived from (14) with a time resolution of 1 day, and $m^{exp}$ is considered to be the discrete version, with the same resolution, of the given from experimental observations mass growth function. The relative error that relation (17) infers is in effect a comparison using the discrete RMS norm.

Besides a small RMS error, the chosen parameter set (16) should satisfy additional conditions of biological nature; for example on the values of the time variables $\tau_h^{1,2}, \tau_r^{1,2}$ in (3) and (4). These additional conditions depend on the species under consideration and will be clarified in each particular experiment.

In practice, for the first part of the evaluation of the parameters (16), that is for the RMS error (17) we employ a parameter estimation technique, which in principle works as follows:

- For a set of parameters (16), evaluate the energy intake and the maintenance and growth energies from (15) and (3)-(5).
- Compute the numerical body mass curve $m^{comp}$ using (14) with $\Delta t = 1$ and compare it with the experimentally observed $m^{exper}$ evaluated the same time instant. The resulting RMS error constitutes the *objective functional*.

- Choose a new set of parameters that decrease the objective functional and repeat the process until a *local minimum* of the objective functional is reached that satisfies (17) and additional species-dependent stopping criteria.

Regarding the third step of the above method, there is plethora of numerical techniques to choose appropriate new parameter sets and according stopping criteria. In this work, we use a combination of the *global optimization method* Enhanced Scatter Search (eSS) (Egea et al., 2009) for the identification of the initial parameter sets for the minimization procedure, augmented with the *local optimization algorithm* Sequential Quadratic Programming (SQP) (Nocedal and Wright, 2006) for a refined local minimization.

The global optimization method we employ belongs to the larger class of *stochastic global optimization methods* called *metaheuristics* (Glover and Kochenberger, 2003). Like other stochastic optimization methods, the eSS draws an initial *diverse* population of guesses out of the parameter space and conditionally initiates intense local searches. The local optimization method SQP is an *iterative nonlinear optimization* method that is particularly well suited for this type of problems due to its robustness in both constraint and un-constrained problems.

For the implementation, we used the Metaheuristics for Bioinformatics Global Optimization (MEIGO) toolbox (Egea et al., 2010) that was developed for the numerical computation environment MATLAB.

# 4    Parameterization of the energy budget model and computation of the energy intake function (Numerical investigations)

For our numerical study, we considered twelve species. These were the fish species Atlantic herring (*Clupea harengus*), and Atlantic cod (*Gadus morhua*), the reptiles green iguana (*Iguana iguana*), Komodo dragon (*Varanus komodoensis*), and American alligator (*Alligator*

*mississippiensis*), the birds blue-and-yellow macaw (*Ara ararauna*), Japanese quail (*Coturnix japonica*), and domestic chicken (*Gallus gallus*), and the mammals black rat (*Rattus rattus*), European hare (*Lepus europaeus*), domestic pig (*Sus scrofa*), and Ayrshire cattle (*Bos taurus*). As described in Section 3, we proceeded as follows: for each of the twelve species we prescribe the experimentally observed body mass growth curve $m^{exp}$, and subsequently identify a full parameter set (16) that a) yields a computational body mass curve $m^{com}$ that satisfies the RMS threshold (17), and b) satisfies additional biologically driven stopping criteria, namely time constraints at which sexual maturity is reached. See the details in the supplementary material given in Section 8.

The body mass growth curves have been experimentally identified to be Bertalanffy curves for the chosen fish and reptile species, logistic curves for chosen birds, and Gompertz curves for chosen mammals.

The Bertalanffy and logistic curves are special cases of the Richards curve:

$$m(t) = M(1 - be^{-kt})^S, \tag{18}$$

where $b = 1 - \left(\frac{m_0}{M}\right)^{\frac{1}{S}}$, and where $m_0$ and $M$ the birth and maximum masses, $k$ the growth rate, and $S$ a species specific parameter ($S = 3$ for the Bertalanffy and $S = -1$ for the logistic). The Gompertz curve is given as

$$m(t) = Me^{-be^{-kt}}, \tag{19}$$

where $b = -\log\left(\frac{m_0}{M}\right)$.

Species-wise, the parameter settings we used for growth curves $m^{exp}$ and longevity have been experimentally identified (Bowling and Putnam, 1943; Case, 1978; Catania, 2016; de Magalhães and Costa, 2009; Gingell, 2016; Hanson, 1987; Lutz and Dunbar-Cooper, 1984; Oosthuizen and Daan, 1974; Pajerski et al., 2016; Pauly, 1980; Ricklefs, 1979; Spiers et al., 1974; Starck and Ricklefs, 1998; Werner and Griebeler, 2014; Zijlstra, 1973; Zullinger et al., 1984) and read as follows:

| Species | growth curve | $m_0$ (g) | $M$ (g) | $k$ (g/day) | longevity in captivity (years), range, mean |
|---|---|---|---|---|---|
| *Clupea harengus* | Bertalanffy | 0.0005 | 350 | 0.000575342 | 30 |
| *Gadus morhua* | Bertalanffy | 0.0000746 | 24000 | 0.000273973 | 65 |
| *Iguana iguana* | Bertalanffy | 11 | 1500 | 0.000315 | 19.8 |
| *Varanus komodoensis* | Bertalanffy | 100 | 63500 | 0.000682759 | 62 (wild) |
| *Alligator mississippiensis* | Bertalanffy | 59.2 | 160000 | 0.000379688 | [60, 85] 75.1 |
| *Coturnix japonica* | Logistic | 4.94 | 113 | 0.106 | 6 |
| *Ara ararauna* | Logistic | 17 | 1116 | 0.126 | 43 |
| *Gallus gallus* | Gompertz | 40 | 5197 | 0.024 | [15, 20] 17.5 |
| *Rattus rattus* | Gompertz | 4.55 | 140 | 0.0207 | 4.2 |
| *Lepus europaeus* | Gompertz | 120 | 4030 | 0.0191 | 10.7 |
| *Sus scrofa* | Gompertz | 960 | 147000 | 0.006 | 27 |
| *Bos taurus (Ayrshire)* | Gompertz | 34470 | 555000 | 0.0032 | 20 |

We note that all the above growth curves are monotonically increasing functions of time. Of particular importance in these curves are the times $t^*$ at which animals reach 99% of their maximum mass, i.e. $m(t^*) = 0.99\,M$. In our numerical computations, we used the times $t^*$ to calibrate the maximum numerical experimentation times, because $M$ is an asymptotic mass (limiting value, $t \to \infty$).

## 5 Results and Discussion (Biological interpretations)

We were able to find, for each of the species we studied, a parameter set (16) that satisfies three major conditions: a) it produces a computational body mass curve $m^{com}$ that satisfies the RMS threshold (17) for $\varepsilon = 0.01$, b) it produces an energy intake versus body mass curve (shown in the lower right graph of each Figure), which is in accordance with the expected for each species, and c) sexual maturity is reached within the expected species dependent ranges (Figure 2-3, see also supplementary material in Section 8 ). This shows that our energy budget model (for graphical representations see Figures 2, 3 upper left panel and in the supplementary Figures 4-15 upper left panel) is able to produce biologically reasonable data for ectothermic and endothermic species without *a priori* informing it on the thermoregulation strategy used by each species. Our results imply consistently with some empirical studies (McClure and Randolph, 1980; Węgrzyn, 2013) that in endothermic species, like birds and mammals, energy is used for heat production instead of growth. Our model additionally offers an explanation on known differences in absolute energy intake between ectothermic fish and reptiles and endothermic birds and mammals. In these ectothermic species absolute energy intake increases linearly with body mass (Pandlan, 1967), whereas in these endotherms it is hyperbolic with an asymptotic value being reached early in their life and long before they are fully grown (Bruggeman et al., 1997; Seebeck et al., 1971; Shindo et al., 2014).

We are aware that due to its generality and particular assumptions, our model will not cover every single species under each situation in nature, especially not when our initial assumptions are violated. Nevertheless, it is sufficiently accurate to describe general patterns of energy intake observed in ectothermic and endothermic vertebrates and offers new insights in the allocation of energy of ectothermic and endothermic species. Our results probably will influence further research in this direction, e.g. the development of more detailed energy budget models for endotherms and ectotherms, and they might also be important in understanding the evolution of endothermy.

a) *Clupea harengus*

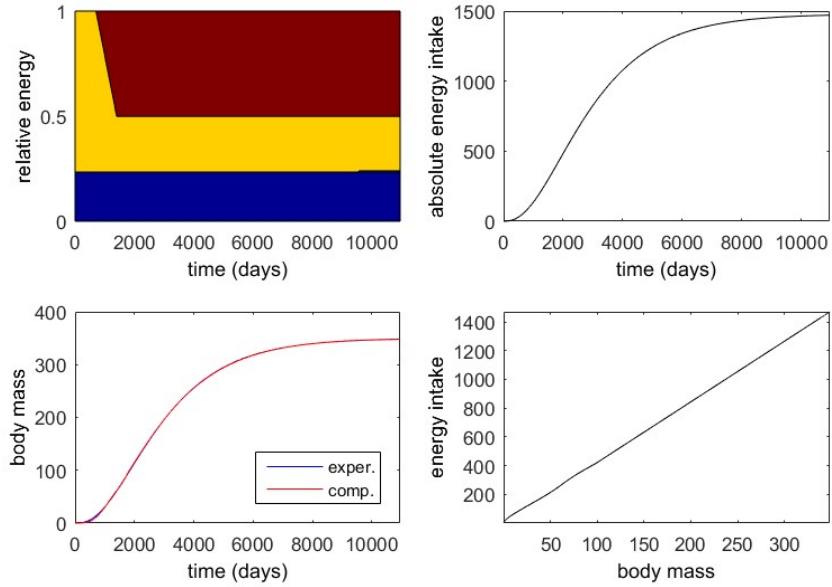

b) *Varanus komodoensis*

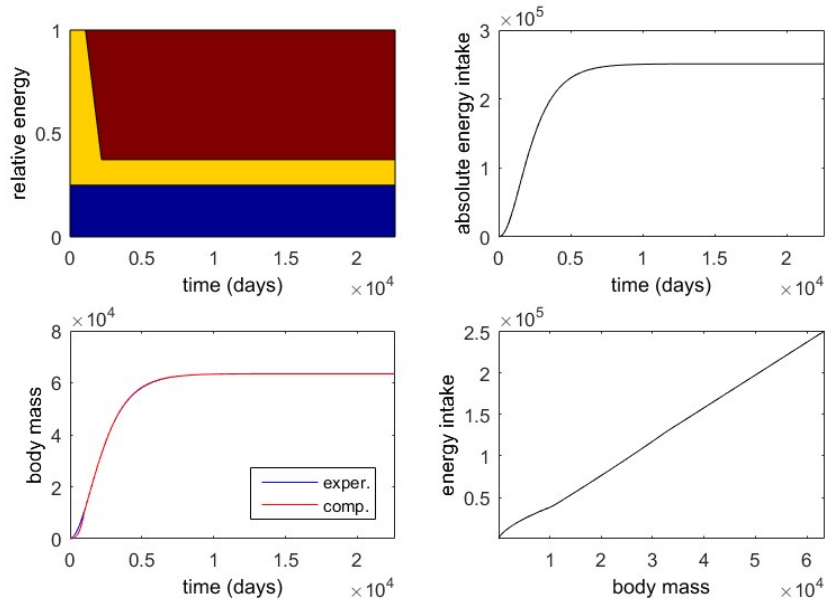

**Figure 2. Ectothermic energy budget models and the resulting total energy intakes and growth trajectories**. The structure of Figures a) and b) is as follows: the *input/given* experimental body mass curve $m^{exp}$ is shown in the lower left figure in blue. The *outputs/results* are shown as: i) the relative energy distribution is shown in the upper left, ii) the absolute energy intake is shown in the upper right, iii) the numerically computed body mass growth $m^{com}$ in the lower left (for comparison with the experimental one), and iv) the energy intake relative to the body mass in the lower right.

a) *Clupea harengus*: We have used a life span of 30 years and sexual maturity of [2,4] years. The output parameters are estimated by the procedures described above in the manuscript and read as follows:
$\{\bar{E}_{main}, \bar{E}_{heat}^{max}, \bar{E}_{repr}^{max}, \tau_h^1, \tau_h^2, \tau_r^1, \tau_r^2\} = \{0.234, 0.006, 0.501, 9548.4, 9548.4, 704, 1401.6\}$
$\{m_0, M, k\} = \{0.056, 1.47\ 10^3, 5.67\ 10^{-4}\}$
$E_{rel} = 0.00255$

b) *Varanus komodoensis*: We have used a life span of 62 years and sexual maturity of [3,6] years. The output parameters are estimated by the procedures described above in the manuscript and read as follows:
$\{\bar{E}_{main}, \bar{E}_{heat}^{max}, \bar{E}_{repr}^{max}, \tau_h^1, \tau_h^2, \tau_r^1, \tau_r^2\} = \{0.252, 4.79\ 10^{-7}, 0.023, 22630, 21951, 1074.9, 2188.3\}$
$\{m_0, M, k\} = \{100, 2.51\ 10^5, 6.99\ 10^{-4}\}$; $E_{rel} = 0.0051$

a) *Gallus gallus*

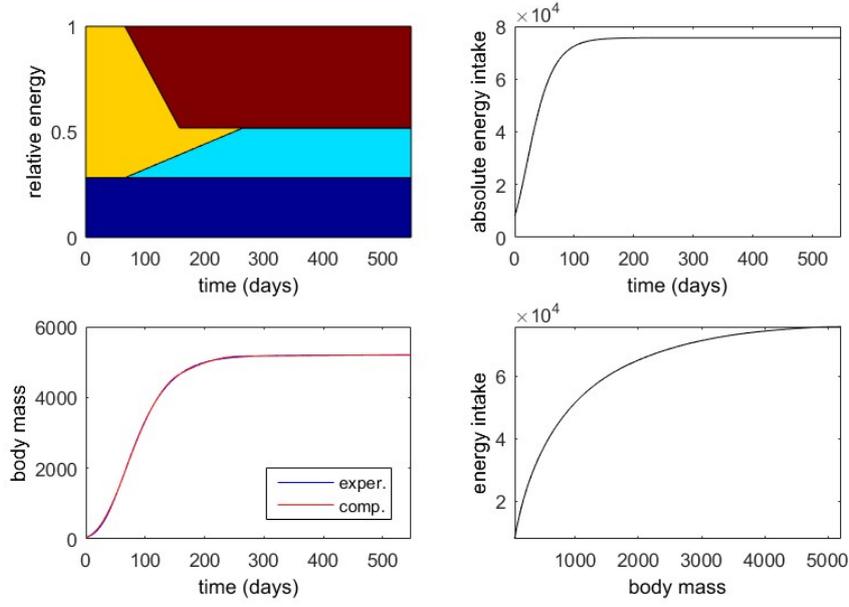

b) *Rattus rattus*

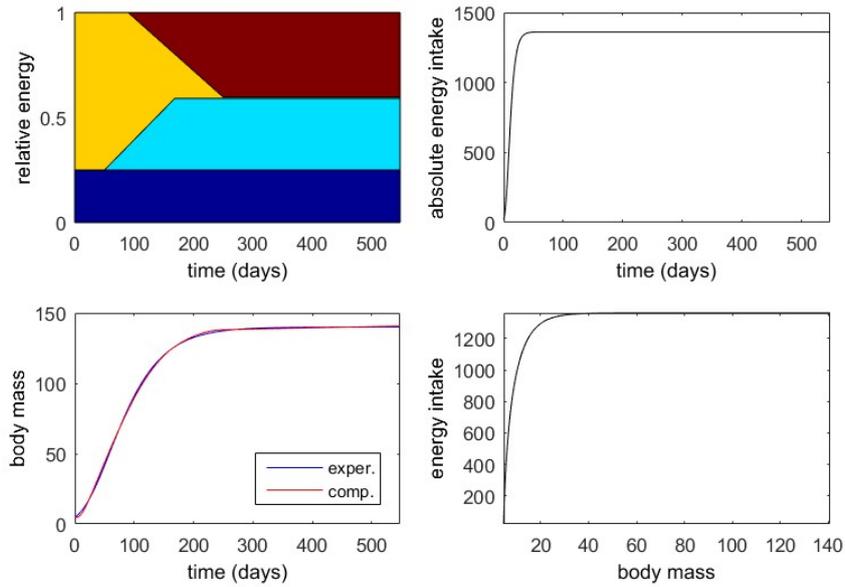

**Figure 3. Endothermic energy budget models and the resulting total energy intakes and growth trajectories**. The structure of Figures a) and b) is as follows: the *input/given* experimental body mass curve $m^{exp}$ is shown in the lower left figure in blue. The *outputs/results* are shown as: i) the relative energy distribution is shown in the upper left, ii) the absolute energy intake is shown in the upper right, iii) the numerically computed body mass growth $m^{com}$ in the lower left (for comparison with the experimental one), and iv) the energy intake relative to the body mass in the lower right.

a) *Gallus gallus*: It has a life span of 17.5 years, we have used 1.5 years for the numerical simulations, and a sexual maturity domain [140, 182.5] days. The output parameters are estimated by the procedures described above in the manuscript and read as follows:
$\{\bar{E}_{main}, \bar{E}_{heat}^{max}, \bar{E}_{repr}^{max}, \tau_h^1, \tau_h^2, \tau_r^1, \tau_r^2\} = \{0.282, 0.234, 0.481, 65.8, 263.2, 65.8, 157.7\}$
$\{m_0, M, k\} = \{8.08 \cdot 10^3, 7.56 \cdot 10^4, 0.039\}$
$$E_{rel} = 0.00317$$

b) *Rattus rattus*: It has a life span of 4.2 years, we have used 1.5 years for the numerical simulations, and a sexual maturity domain [1,2] years. The output parameters are estimated by the procedures described above in the manuscript and read as follows:
$\{\bar{E}_{main}, \bar{E}_{heat}^{max}, \bar{E}_{repr}^{max}, \tau_h^1, \tau_h^2, \tau_r^1, \tau_r^2\} = \{0.251, 0.341, 0.402, 49.8, 168.0, 89.7, 249.1\}$
$\{m_0, M, k\} = \{21, 1.36 \cdot 10^3, 0.165\}$; $E_{rel} = 0.00701$

# 6 Acknowledgements

JW was partially financed by the German Research Foundation (grant GR 2625/2-2). NS was partially financed by the Alexander von Humboldt-Foundation. JW and NS were partially financed by the Johannes Gutenberg University of Mainz Center for Computational Sciences (SRFN).

# 7 References

Bowling, G.A., and Putnam, D.N., 1943. Growth Studies with Ayrshire Cattle. I. Normal Body Weights and Heights at Shoulders for Ayrshire Cattle1. J Dairy Sci 26, 47-52.

Brown, J.H., Gillooly, J.F., Allen, A.P., Savage, V.M., and West, G.B., 2004. Toward a metabolic theory of ecology. Ecology 85, 1771-1789.

Bruggeman, V., Vanmontfort, D., Renaville, R., Portetelle, D., and Decuypere, E., 1997. The Effect of Food Intake from Two Weeks of Age to Sexual Maturity on Plasma Growth Hormone, Insulin-like Growth Factor-I, Insulin-like Growth Factor-Binding Proteins, and Thyroid Hormones in Female Broiler Breeder Chickens. Gen Comp Endocrinol 107, 212-220.

Case, T.J., 1978. On the evolution and adaptive significance of postnatal-growth rates in terrestrial vertebrates. Q Rev Biol 53, 243-282.

Catania, K., Ara ararauna blue-and-yellow macaw. Animal Diversity Web: http://animaldiversity.org/accounts/Ara_ararauna/ accessed on 08.04.2016, 2016.

Clarke, A., and Pörtner, H.-O., 2010. Temperature, metabolic power and the evolution of endothermy. Biol Rev 85, 703-727.

Cowles, R.B., 1940. Additional implications of reptilian sensitivity to high temperatures. Am Nat 74, 542-561.

de Magalhães, J.P., and Costa, J., 2009. A database of vertebrate longevity records and their relation to other life-history traits. J Evol Biol 22, 1770-1774.

Egea, J.A., Martí, R., and Banga, J.R., 2010. An evolutionary method for complex-process optimization. Computers & Operations Research 37, 315-324.

Egea, J.A., Balsa-Canto, E., García, M.-S.G., and Banga, J.R., 2009. Dynamic Optimization of Nonlinear Processes with an Enhanced Scatter Search Method. Industrial & Engineering Chemistry Research 48, 4388-4401.

Gingell, F., Iguana iguana Common Green Iguana, Animal Diversity Web: http://animaldiversity.org/accounts/Iguana_iguana/, accessed on 08.04.2016, 2016.

Glover, F., and Kochenberger, G.A., 2003. Handbook of Metaheuristics. Springer US, New York.

Grigg, G.C., Beard, L.A., and Augee, M.L., 2004. The evolution of endothermy and its diversity in mammals and birds. Physiol Biochem Zool 77, 982-997.

Hanson, J.T., 1987. Handraising large parrots: Methodology and expected weight gains. Zoo Biol 6, 139-160.

Hou, C., Zuo, W., Moses, M.E., Woodruff, W.H., Brown, J.H., and West, G.B., 2008. Energy Uptake and Allocation During Ontogeny. Science 322, 736-739.

Hulbert, A.J., and Else, P.L., 1981. Comparison of the "mammal machine" and the "reptile machine": energy use and thyroid activity. AJP - Regulatory, Integrative and Comparative Physiology 241, R350-356.

Kooijman, S.A.L.M., 2010. Dynamic energy budget theory for metabolic organisation. Cambridge University Press.

Koteja, P., 2004. The evolution of concepts on the evolution of endothermy in birds and mammals. Physiol Biochem Zool 77, 1043–1050.

Kozlowski, J., 1992. Optimal allocation of resources to growth and reproduction: implications for age and size at maturity. Trends Ecol Evol 7, 15-19.

Kozlowski, J., 1996. Optimal Allocation of Resources Explains Interspecific Life-History Patterns in Animals with Indeterminate Growth. Proceedings: Biological Sciences 263, 559-566.

Kozlowski, J., and Weiner, J., 1997. Interspecific Allometries Are by-Products of Body Size Optimization. Am Nat 149, 352-380.

Lutz, P.L., and Dunbar-Cooper, A., 1984. The Nest Environment of the American Crocodile (Crocodylus acutus). Copeia 1984, 153-161.

McClure, P.A., and Randolph, J.C., 1980. Relative Allocation of Energy to Growth and Development of Homeothermy in the Eastern Wood Rat (Neotoma floridana) and Hispid Cotton Rat (Sigmodon hispidus). Ecol Monogr 50, 199-219.

McNab, B.K., 1978. The evolution of endothermy in the phylogeny of mammals. Am Nat 112, 1-21.

Nespolo, R.F., Bacigalupe, L.D., Figueroa, C.C., Koteja, P., and Opazo, J.C., 2011. Using new tools to solve an old problem: the evolution of endothermy in vertebrates. Trends Ecol Evol 26, 414-423.

Nocedal, J., and Wright, S., 2006. Numerical Optimization. Springer, New York.

Oosthuizen, E., and Daan, N., 1974. Egg fecundity and maturity of North Sea cod, gadus morhua. Netherlands Journal of Sea Research 8, 378-397.

Pajerski, L., Schechter, B., and Street, R., Alligator mississippiensis Alligator, Gator, American alligator, Florida alligator, Mississippi alligator, Louisiana alligator. Animal Diversity Web: http://animaldiversity.org/accounts/Alligator_mississippiensis/ accessed on 08.04.2016, 2016.

Pandlan, T.J., 1967. Intake, digestion, absorption and conversion of food in the fishes Megalops cyprinoides and Ophiocephalus striatus. Mar Biol 1, 16-32.

Pauly, D., 1980. On the interrelationships between natural mortality, growth parameters, and mean environmental temperature in 175 fish stocks. Journal du Conseil 39, 175-192.

Ricklefs, R.E., 1979. Patterns of growth in birds. V. A comparative study of development in the Starling, Common Tern, and Japanese Quail. Auk 96, 10-30.

Ricklefs, R.E., 1987. Characterizing the development of homeothermy by rate of body cooling. Funct Ecol 1, 151-157.

Ruben, J., 1995. The evolution of endothermy in mammals and birds: from physiology to fossils. Annu Rev Physiol 57, 69-95.

Schweitzer, M.H., and Marshall, C.L., 2001. A molecular model for the evolution of endothermy in the theropod-bird lineage. J Exp Zool 291, 317-338.

Seebeck, R., Springell, P., and O'kelly, J., 1971. Alterations in Host Metabolism by the Specific and Anorectic Effects of the Cattle Tick (Boophilus Microplus) I. Food Intake and Body Weight Growth. Australian Journal of Biological Sciences 24, 373-380.

Shindo, D., Matsuura, T., and Suzuki, M., 2014. Effects of prepubertal-onset exercise on body weight changes up to middle age in rats. J Appl Physiol 116, 674-682.

Sousa, T., Domingos, T., and Kooijman, S.A.L.M., 2008. From empirical patterns to theory: a formal metabolic theory of life. Philosophical Transactions of the Royal Society B-Biological Sciences 363, 2453-2464.

Spiers, D., McNabb, R., and McNabb, F.M.A., 1974. The development of thermoregulatory ability, heat seeking activities, and thyroid function in hatchling Japanese quail (Coturnix coturnix japonica). Journal of comparative physiology 89, 159-174.

Starck, J.M., and Ricklefs, R.E. Eds.), 1998. Avian growth and development. Evolution within the altricial-precocial spectrum. Oxford University Press, New York.

Visser, G.H., and Ricklefs, R.E., 1995. Relationship between Body Composition and Homeothermy in Neonates of Precocial and Semiprecocial Birds. The Auk 112, 192-200.

Węgrzyn, E., 2013. Resource allocation between growth and endothermy allows rapid nestling development at low feeding rates in a species under high nest predation. J Avian Biol 44, 383-389.

Werner, J., and Griebeler, E.M., 2014. Allometries of maximum growth rate versus body mass at maximum growth indicate that non-avian dinosaurs had growth rates typical of fast growing ectothermic sauropsids. PLoS ONE 9, e88834.

West, G.B., Brown, J.H., and Enquist, B.J., 2001. A general model for ontogenetic growth. Nature 413, 628-631.

White, C.R., Phillips, N.F., and Seymour, R.S., 2006. The scaling and temperature dependence of vertebrate metabolism. Biol Lett 2, 125-127.

Zijlstra, J.J., 1973. Egg weight and fecundity in the north sea herring (clupea harengus). Netherlands Journal of Sea Research 6, 173-204.

Zullinger, E.M., Ricklefs, R.E., Redford, K.H., and Mace, G.M., 1984. Fitting sigmoidal equations to mammalian growth curves. J Mammal 65, 607-636.

# 8 Supplementary material

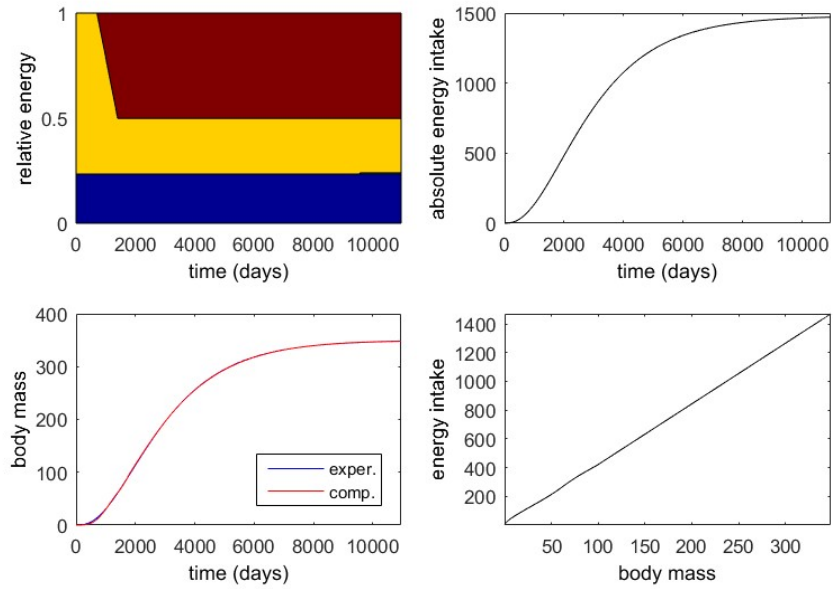

**Figure 4.** *Clupea harengus*: We have used a life span of 30 years and sexual maturity of [2,4] years.

$\{\bar{E}_{main}, \bar{E}_{hea}^{max}, \bar{E}_{repr}^{max}, \tau_h^1, \tau_h^2, \tau_r^1, \tau_r^2\} = \{0.234, 0.006, 0.501, 9548.4, 9548.4, 704, 1401.6\}$
$\{m_0, M, k\} = \{0.056, 1.47 \ 10^3, 5.67 \ 10^{-4}\}$
$E_{rel} = 0.00255$

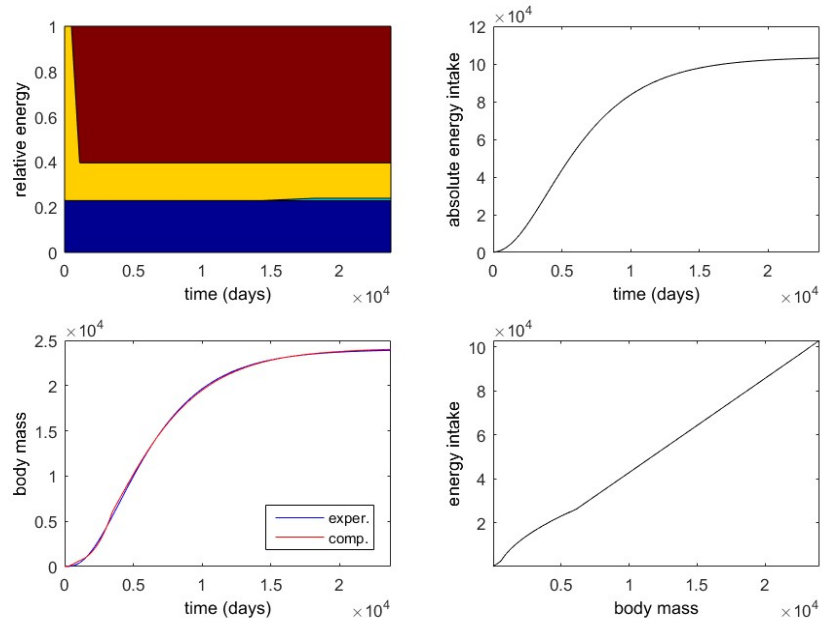

**Figure 5.** *Gadus morhua*: We have used a life span of 65 years and sexual maturity of [2,3] years.

$\{\bar{E}_{main}, \bar{E}_{hea}^{max}, \bar{E}_{repr}^{max}, \tau_h^1, \tau_h^2, \tau_r^1, \tau_r^2\} = \{0.229, 0.01, 0.604, 14330, 18102, 4935, 1093.7\}$
$\{m_0, M, k\} = \{99, 1.035 \ 10^5, 2.56 \ 10^{-4}\}$
$E_{rel} = 0.00761$

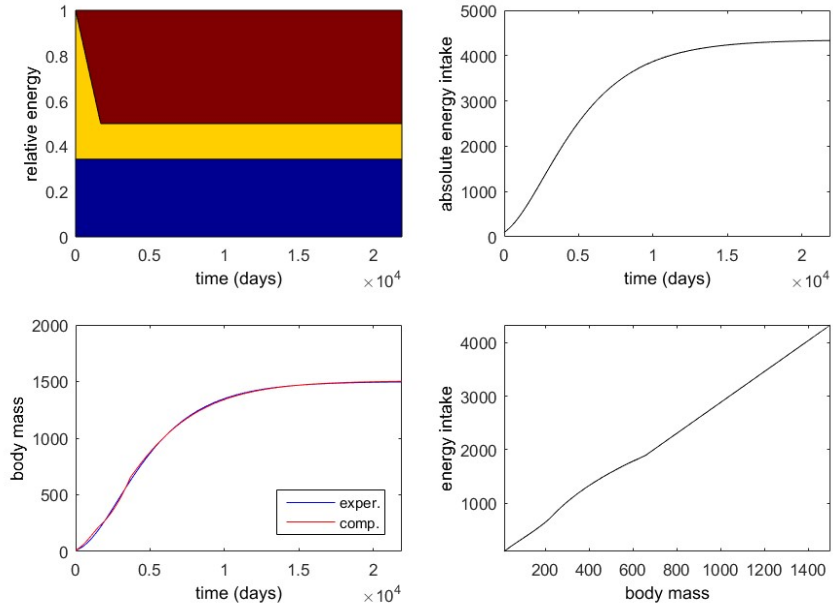

**Figure 6.** *Iguana Iguana*: We have used a life span of 60 years and sexual maturity of [2.5,5] years.

$\{\bar{E}_{main}, \bar{E}_{heat}^{max}, \bar{E}_{repr}^{max}, \tau_h^1, \tau_h^2, \tau_r^1, \tau_r^2\} = \{0.343, 0, 0.5, 79, 836.58, 0, 1684.1\}$
$\{m_0, M, k\} = \{100, 4.35\ 10^3, 2.93\ 10^{-4}\}$
$E_{rel} = 0.00781$

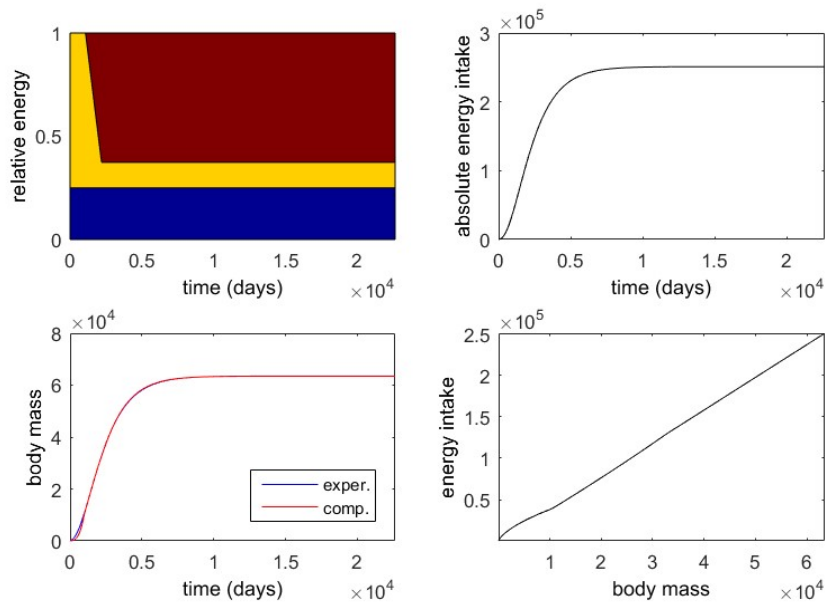

**Figure 7.** *Varanus komodoensis*: We have used a life span of 62 years and sexual maturity of [3,6] years.

$\{\bar{E}_{main}, \bar{E}_{heat}^{max}, \bar{E}_{repr}^{max}, \tau_h^1, \tau_h^2, \tau_r^1, \tau_r^2\} = \{0.252, 4.79\ 10^{-7}, 0.023, 22630, 21951, 1074.9, 2188.3\}$
$\{m_0, M, k\} = \{100, 2.51\ 10^5, 6.99\ 10^{-4}\}$
$E_{rel} = 0.0051$

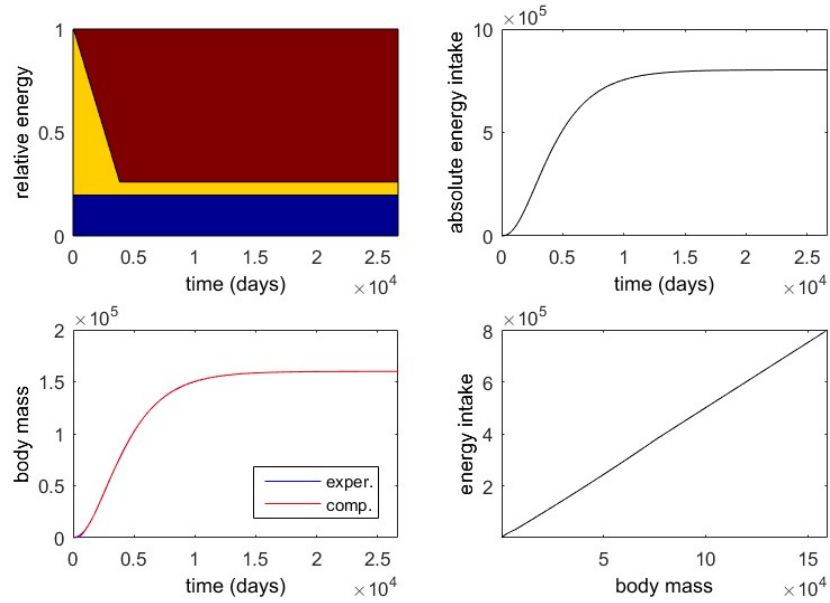

**Figure 8.** *Alligator mississippiensis*: We have used a life span of 73.1 years and sexual maturity of [10,12] years.

$\{\bar{E}_{main}, \bar{E}_{hea}^{max}, \bar{E}_{repr}^{max}, \tau_h^1, \tau_h^2, \tau_r^1, \tau_r^2\} = \{0.197, 0, 0.738, 26681, 26681, 50.6, 3815.5\}$
$\{m_0, M, k\} = \{78, 8.02\ 10^5, 3.84\ 10^{-4}\}$
$E_{rel} = 0.00144$

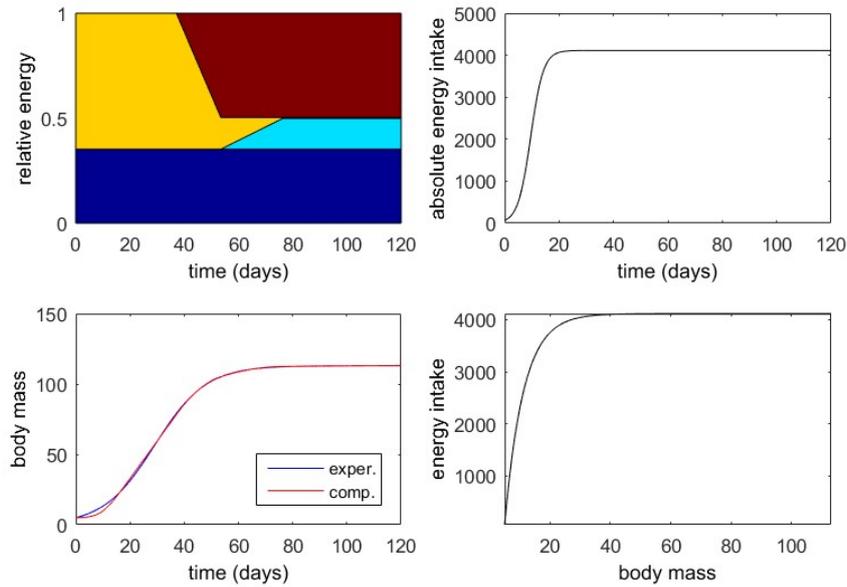

**Figure 9.** *Coturnix japonica*: We have used a domain of 120 days for the computation, and assume a life span of 3 years, and sexual maturity of [52,63] days.

$\{\bar{E}_{main}, \bar{E}_{heat}^{max}, \bar{E}_{repr}^{max}, \tau_h^1, \tau_h^2, \tau_r^1, \tau_r^2\} = \{0.352, 0.146, 0.496, 53.4, 76.3, 37.0, 53.5\}$
$\{m_0, M, k\} = \{66, 4.11\ 10^3, 0.428\}$
$E_{rel} = 0.01094$

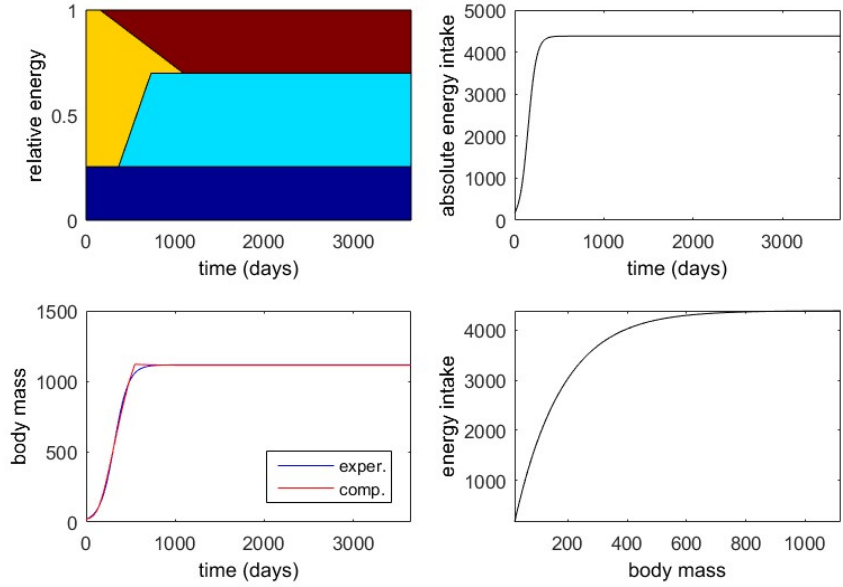

**Figure 10.** *Ara ararauna*: It has a life span of 43 years, we have used 10 years for the numerical simulations, and a sexual maturity domain [3,4] years.

$\{\bar{E}_{main}, \bar{E}_{heat}^{max}, \bar{E}_{repr}^{max}, \tau_h^1, \tau_h^2, \tau_r^1, \tau_r^2\} = \{0.254, 0.445, 0.300, 365, 730, 158, 6, 1095\}$
$\{m_0, M, k\} = \{175, 4.3 \cdot 10^3, 0.0021\}$
$E_{rel} = 0.00991$

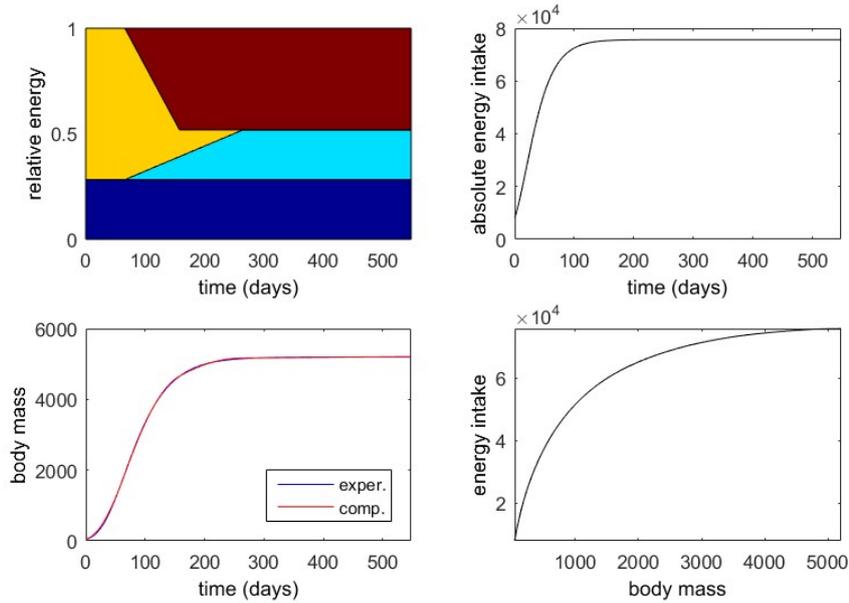

**Figure 11.** *Gallus gallus*: It has a life span of 17.5 years, we have used 1.5 years for the numerical simulations, and a sexual maturity domain [140,182.5] days.

$\{\bar{E}_{main}, \bar{E}_{hea}^{max}, \bar{E}_{repr}^{max}, \tau_h^1, \tau_h^2, \tau_r^1, \tau_r^2\} = \{0.282, 0.234, 0.481, 65.8, 263.2, 65.8, 157.7\}$
$\{m_0, M, k\} = \{8.08 \cdot 10^3, 7.56 \cdot 10^4, 0.039\}$
$E_{rel} = 0.00317$

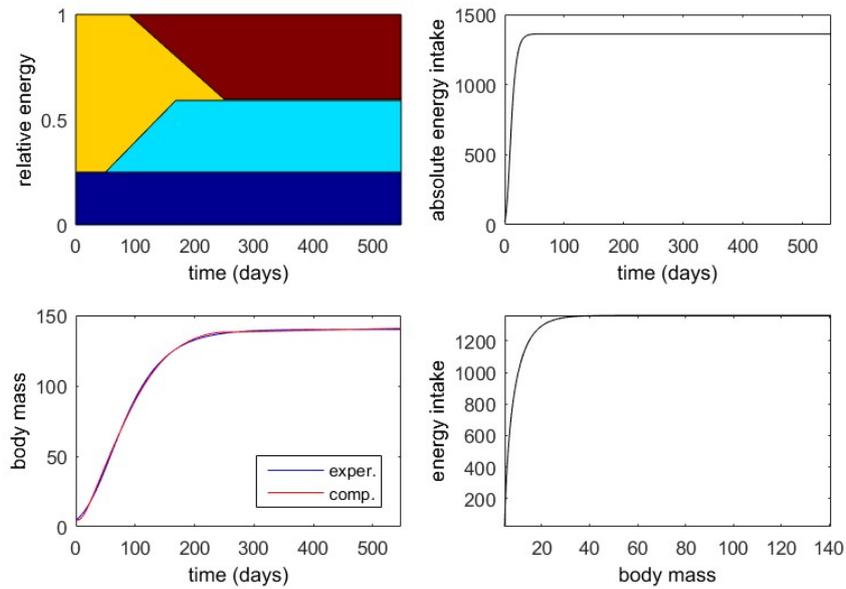

**Figure 12.** *Rattus rattus*: It has a life span of 4.2 years, we have used 1.5 years for the numerical simulations, and a sexual maturity domain [1,2] years.

$\{\bar{E}_{main}, \bar{E}_{heat}^{max}, \bar{E}_{repr}^{max}, \tau_h^1, \tau_h^2, \tau_r^1, \tau_r^2\} = \{0.251, 0.341, 0.402, 49.8, 168.0, 89.7, 249.1\}$
$\{m_0, M, k\} = \{21, 1.36\ 10^3, 0.165\}$
$E_{rel} = 0.00701$

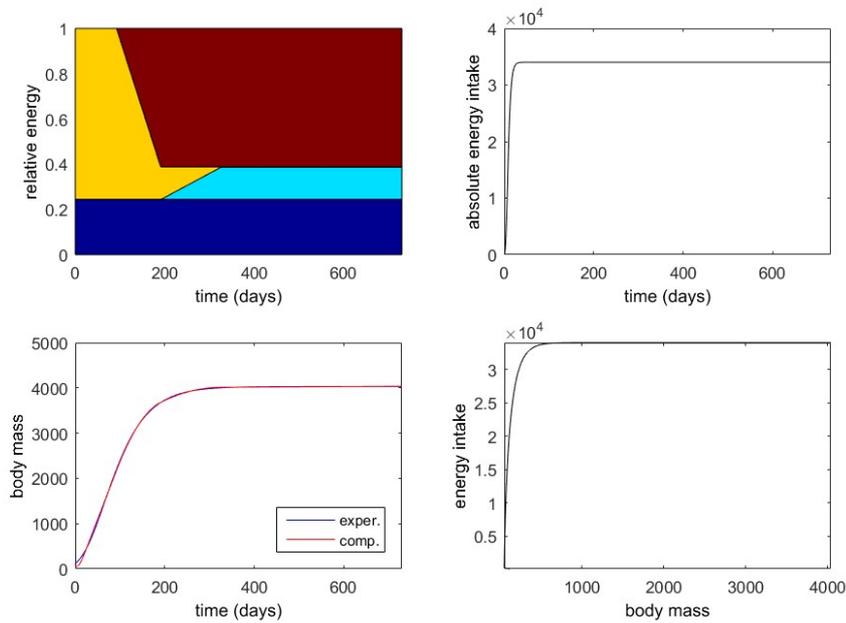

**Figure 13.** *Lepus europaeus*: It has a life span of 10.7 years, we have used 2 years for the numerical simulations, and a sexual maturity domain [0.5,1] years.

$\{\bar{E}_{main}, \bar{E}_{heat}^{max}, \bar{E}_{repr}^{max}, \tau_h^1, \tau_h^2, \tau_r^1, \tau_r^2\} = \{0.244, 0.141, 0.611, 191.9, 325.5, 91.9, 190.5\}$
$\{m_0, M, k\} = \{78.9, 3.40\ 10^4,\ 0.243\}$
$E_{rel} = 0.00551$

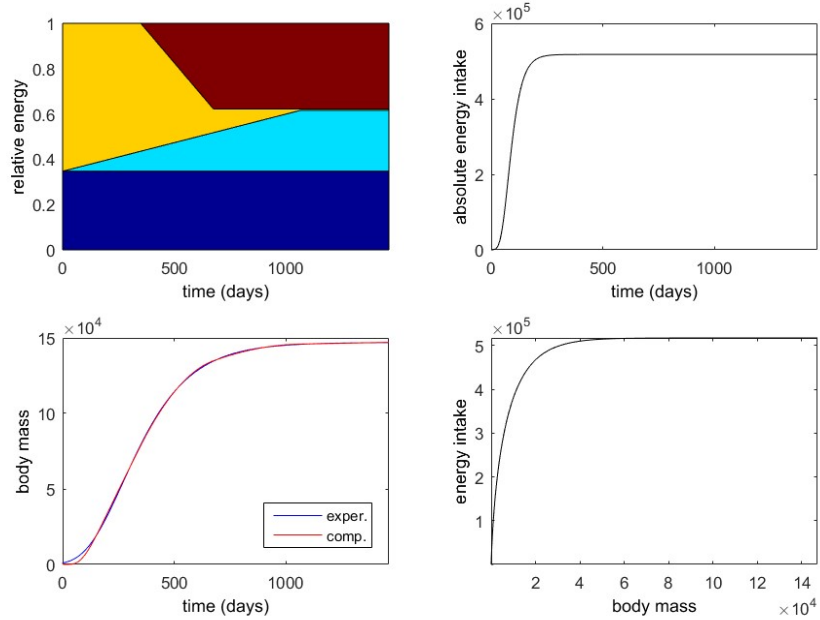

**Figure 14.** *Sus scrofa*: It has a life span of 27 years, we have used 4 years for the numerical simulations, and a sexual maturity domain [334,768] days.

$\{\bar{E}_{main}, \bar{E}_{heat}^{max}, \bar{E}_{repr}^{max}, \tau_h^1, \tau_h^2, \tau_r^1, \tau_r^2\} = \{0.34, 0.269, 0.378, 0.1, 1066, 349.4, 674.8\}$
$\{m_0, M, k\} = \{38, 5.17\ 10^5, 0.029\}$
$E_{rel} = 0.00761$

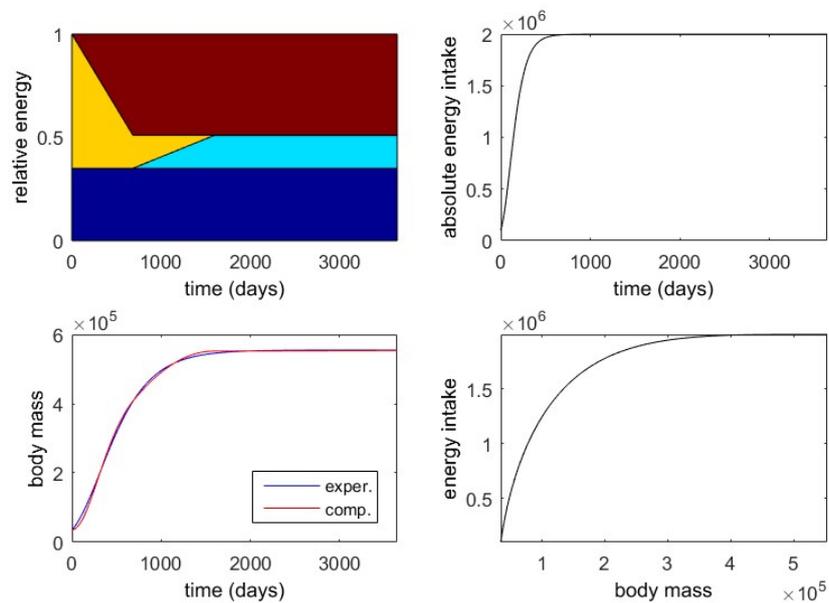

**Figure 15.** *Bos taurus*: It has a life span of 20 years, we have used 10 years for the numerical simulations, and a sexual maturity domain [1,2] years.

$\{\bar{E}_{main}, \bar{E}_{heat}^{max}, \bar{E}_{repr}^{max}, \tau_h^1, \tau_h^2, \tau_r^1, \tau_r^2\} = \{0.35, 0.159, 0.489, 681.5, 1596.7, 0, 684.3\}$
$\{m_0, M, k\} = \{10^5, 1.9\ 10^6, 0.01\}\{Węgrzyn, 2013\ \#2128\}$
$E_{rel} = 0.00708$

# 9 A discrete approach to (8)

The energy contained within an animal during the $n$-th day of its life, i.e. during the time period $[t^{n-1}, t^n]$, $n = 1, \cdots$, is given by two parts: one part that accounts for the newly invested growth energy which accumulates with the previous growth-invested energy, and another part for the energy invested for maintenance. Namely:

$$E_{maint}^{[n-1,n]} + E_{grow}^{[n-1,n]} + E_{grow}^{\leq n-1}. \tag{20}$$

Motivated by the circadian rhythm and the daily patterns of the animal, we consider the time period of *one day* (and not fraction of it) as the *unit of time*. We moreover assume that the energy intake and the energy investment in maintenance and growth (denoted discretely as $E_{intake}^n$, $E_{maint}^n$, and $E_{grow}^n$) follow uniform distribution during each day.

Setting $\Delta t$ to represent the time period of one day, (20) reads after an iterative $n$ days cycle as:

$$\Delta t\, E_{maint}^n + \Delta t \sum_{k=1}^{n-1} E_{grow}^k + \Delta t\, E_{grow}^n. \tag{21}$$

The energy content that (21) describes can be "translated" to (total) body mass via a *mass-to-energy* transition coefficient $C_m$. Namely, if $m^n$ is the body mass at the $n$-th day of the animals life, it holds

$$C_m m^n = \Delta t\, E_{maint}^n + \Delta t \sum_{k=1}^{n-1} E_{grow}^k + \Delta t\, E_{grow}^n. \tag{22}$$

We consider at this point the energy terms such as $E_{maint}^n, E_{grow}^n$ as time rates of energy investments with units $\left[\frac{J}{d}\right]$ (joules per day), and the mass-to-energy coefficient $C_m$ with units $\left[\frac{J}{g}\right]$ (joules per gram). In effect the *dimensional analysis* of (22) reads (with the notation $\{U\}$ and $[U]$ for the *numerical value* and the *units* respectively of a *quantity U*, i.e. $U = \{U\}[U]$) as

$$\{C_m\}\left[\frac{J}{g}\right]\{m^n\}[g]$$
$$= \{\Delta t\}[d]\{E_{maint}^n\}\left[\frac{J}{d}\right] + \{\Delta t\}[d]\sum_{k=1}^{n-1}\{E_{grow}^k\}\left[\frac{J}{d}\right] \tag{23}$$
$$+ \{\Delta t\}[d]\{E_{grow}^n\}\left[\frac{J}{d}\right]$$

or

$$\{C_m\}\{m^n\}[J] = \{\Delta t\}\{E_{maint}^n\}[J] + \{\Delta t\}\sum_{k=1}^{n-1}\{E_{grow}^k\}[J] + \{\Delta t\}\{E_{grow}^n\}[J]. \tag{24}$$

After discarding the units $[J]$, replacing the numerical values $\{\cdot\}$ of the variables by their regular notation, and after employing the energy distribution assumptions (2), (24) reads as

$$m^n = \frac{1}{C_m}\Delta t\, \bar{E}_{maint} E_{intake}^n + \frac{1}{C_m}\Delta t \sum_{k=1}^{n} \bar{E}_{grow}^k E_{intake}^k. \tag{25}$$

Finally, we absorb the constant $\frac{1}{C_m}$ inside $E_{intake}$, and take the *formal limit* as the number of days of life $n$ increases. In this perspective the sum in the right-hand side can be viewed as an integral over the continuous time variable $t$, and in effect deduce (J. A. Egea, 2009) the dimensionless version (8) of the model.